# Electrical control over single hole spins in nanowire quantum dots

V. S. Pribiag[1*], S. Nadj-Perge[1], S. M. Frolov[1], J. W. G. van den Berg[1], I. van Weperen[1], S. R. Plissard[2],

E. P. A. M. Bakkers[1,2] and L. P. Kouwenhoven[1*]

1. Kavli Institute of Nanoscience, Delft University of Technology, 2600 GA Delft, The Netherlands
2. Department of Applied Physics, Eindhoven University of Technology, 5600 MB Eindhoven, The Netherlands

**Single electron spins in semiconductor quantum dots (QDs) are a versatile platform for quantum information processing, however controlling decoherence remains a considerable challenge.[1-4] Recently, hole spins have emerged as a promising alternative. Holes in III-V semiconductors have unique properties, such as strong spin-orbit interaction and weak coupling to nuclear spins, and therefore have potential for enhanced spin control[5-8] and longer coherence times[8-12]. Weaker hyperfine interaction has already been reported in self-assembled quantum dots using quantum optics techniques.[10-12] However, challenging fabrication has so far kept the promise of hole-spin-based electronic devices out of reach in conventional III-V heterostructures.[13] Here, we report gate-tuneable hole quantum dots formed in InSb nanowires. Using these devices we demonstrate Pauli spin blockade and electrical control of single hole spins. The devices are fully tuneable between hole and electron QDs, enabling direct comparison between the hyperfine interaction strengths, g-factors and spin blockade anisotropies in the two regimes.**

The development of viable quantum computation devices depends on the ability to preserve the coherence of quantum bits (qubits).[1] To date, most work in semiconductors has focused on using the spins of electrons as qubits. However, due to strong hyperfine coupling to the surrounding nuclear spin bath, electron spins experience rapid decoherence.[2-4] In contrast, the contact hyperfine interaction for hole spins is predicted to vanish due to the p-orbital symmetry of their Bloch wavefunction, leaving only the weaker dipole-dipole coupling.[9] However, a major technical challenge for investigating hole spins in gate-defined quantum dots has been that conventional p-doped quantum wells do not provide the required level of stability and tuneability.[13] An alternative approach is to rely on an undoped structure with a small enough bandgap that the hole transport regime can be accessed simply by means of gate tuning. InSb is an excellent candidate, having the smallest bandgap in the III-V family of semiconductors.[14] InSb nanowires can be grown defect-free using bottom-up methods that offer precise control over crystal structure and electronic properties.[15-17] In addition to their small bandgap, these nanowires are of high interest for quantum devices due to the presence of strong spin-orbit coupling and large g-factors, which enabled the detection of signatures of Majorana fermions in the solid state[18] and allow fast all-electrical spin control[4,19].

To illustrate tuning between electron and hole transport, we first use a basic device consisting of an InSb nanowire placed above a doped Si backgate (Fig. 1). For positive backgate voltages, $V_{BG}$, the Fermi level is tuned into the conduction band, resulting in a source-drain current of ~1μA due to accumulation of electrons. The current is suppressed as $V_{BG}$ is decreased to around -5 V and the Fermi level enters the bandgap. At around -14 V, current is then restored due to accumulation of holes.[16] We rely on this basic tuneability for defining hole quantum dots in the nanowires.

To confine holes to quantum dots, we fabricated devices with five narrow gates that allow local control of the Fermi level (Fig. 2a).[15,19] Taking advantage of the small bandgap of InSb, a hole QD can be formed by inducing holes into a short section of the nanowire using a single fine gate (gate 2). The rest of the nanowire remains n-type, as shown schematically in Fig. 2b. Strong band bending above gate 2 forms an attractive potential well for holes between two p-n junctions.[20] A transition from an entirely n-type wire to such an npn QD is displayed in Fig. 2c for device d1. As the gate-2 voltage, $V_G$, is decreased from around zero, the current through the n-type device is suppressed by the formation of a tunnel barrier above the gate (see right schematic in Fig. 2c). As $V_G$ is decreased further, we observe Coulomb diamonds, indicating that transport in this region is mediated by the first few hole states of the QD (see left schematic in Fig. 2c). Interestingly, the 0-hole Coulomb diamond crosses the boundary of the electron transport regime at a bias of ~90 mV, which corresponds to less than half the value of the bandgap for our InSb nanowires (~200 meV, see Supplementary Fig. S1). This suggests that at larger bias (or at less negative $V_G$), electrons tunnel more efficiently directly through the conduction band barrier than through the two p-n junctions adjacent to the QD (see tunnelling schematics in Fig. 2c). A more detailed view of the first 5 hole diamonds (Fig. 2d), shows distinct resonances in the QD transport regime. These resonances are typical signatures of QD excited states[3], but can also originate from singularities in the density of states in the leads[21]. We find that for our hole QDs the two mechanisms can be distinguished based on their different magnetic field dependence (see Supplementary Fig. S3).

To investigate spin-dependent transport for holes we rely on double quantum dots tuned to the Pauli spin blockade[3] regime. Pauli spin blockade provides a reliable means to initialize and read out spin states in a double dot, and has been studied extensively for the case of electrons.[3] Whenever the spins of unpaired electrons on separate dots form a triplet state, tunnelling between the dots is suppressed, since the final state with the two electrons on the same orbital must be a singlet. Holes in III-V semiconductors differ in many respects from electrons, being characterized by different effective masses and strong coupling of spin to p-orbitals.[22] To form hole double quantum dots we use either gates 2 and 4 (achieving weak interdot coupling) or gates 2 and 3 (achieving strong interdot coupling) to confine holes in adjacent segments of the nanowire (Fig. 3a). Just as in the case of single dots, tunnelling between a double dot and the leads occurs through p-n junctions. The two gates can be used to tune between an entirely n-type wire and an nppn double dot, as shown in the inset of Fig. 3b. More generally, the devices can be fully tuned between hole double dots and electron double dots by varying the voltages on all five fine gates. The electron regime has been studied in detail in Ref. [19] using device d2. The main panel of Fig. 3b shows a zoomed-in view of the charge stability diagram of a few-hole double dot for device d2. Triple-point bias triangles separate regions where the number of holes in each dot is fixed by Coulomb blockade. Alternating spacing between the triple point transitions, together with the absence of lower charge states in the stability diagram, suggest that the double dot occupancy can be controlled down to the last hole (however, charge sensing would be necessary to unambiguously determine the occupancy). Charging energies, $E_c$, are ~20 meV, while the orbital separation, $E_{orb}$, extracted from excited state measurements on the (1,0) state is ~8 meV, in agreement with the single-dot values. It is useful to note that the ratio $E_c^2/E_{orb}$ is proportional to the effective mass, but is independent of the QD size in a first approximation.[3] By comparing the value of $E_c^2/E_{orb}$ obtained here for holes with the value for electrons[19], we find that the effective hole mass in our QDs is comparable to that of electrons, in agreement with expectations for light holes (LH) in InSb ($m_{LH} \approx 0.015\, m_e$, $m_e^* \approx 0.014\, m_e$), and more than an order of magnitude lighter than the heavy hole (HH) mass ($m_{HH} \approx 0.43 m_e$).[14] This suggests that the states in our double dots are predominantly LH-like. Moreover, light holes are also likely to dominate charge transport due to the angular momentum mismatch between heavy holes ($J_z = 3/2$) and spin-1/2 electrons in the leads.

Spin blockade is lifted by any process that mixes triplets and singlets, such as single-spin rotations. In the presence of strong spin-orbit interaction (SOI), spin rotations can be induced by a microwave-frequency electric field applied with one of the plunger gates. This electric dipole spin resonance (EDSR)[23,24] mechanism is expected to drive single hole spin rotations whenever the frequency, $f_0$, of the applied electric field matches the Zeeman energy of the hole spin ($f_0 = g\mu_B B/h$, where $g$ is the Landé g-factor, $\mu_B$ is the Bohr magneton, $B$ is the applied static magnetic field and $h$ is Planck's constant). Lifting of the spin blockade by EDSR allows holes to tunnel between the two dots (Fig. 4a).

Fig. 4b shows the V-shaped resonance, mapped out by measuring the double-dot current as a function of $f_0$ and $B$. In addition to the EDSR lines, we also observe a current peak near $B = 0$. By analogy with the case of electrons, we attribute this feature to the hyperfine interaction (see also Supplementary Fig. S4 for more details about hole spin blockade in the case of weak interdot coupling).[3,25-27] From the peak width we estimate the RMS fluctuations of the hyperfine interaction for holes, $E_{N,h}$ ~0.8 μeV, using the standard method described in Refs. [28] and [29] (see Supplementary Information). Thanks to the extensive tuneability, we are able to perform the same analysis on data from the electron double dot regime of the same device, obtaining $E_{N,e}$ ~2 μeV. Taking into account the electron and hole dot volumes, we estimate the ratio of the hyperfine coupling strength for electrons to that of holes, $A_e/A_h$ ~7. This value is close to the values obtained by optical techniques on self-assembled dots.[30]

The hole g-factor extracted from EDSR spectra (Fig. 4c), shows remarkable differences from the electron g-factor[19] in both magnitude and anisotropy. The magnitude ranges between ~0.5 and ~4 for holes, far less than the typical range of ~35 to ~45 for electrons. The hole g-factor is expected to be a strong function of the nanowire subband[31], ranging between about 0 and several times the value of the bulk hole g-factor $\kappa$ ($\kappa = 15.6$ for InSb [22]). The small g-factor values and strong anisotropy we observe could therefore be a consequence of hole mixing[25,26] due to confinement by the QD potential. An important point, evident when comparing the data for the strong and the weak coupling regimes, is that both the magnitude and the anisotropy of the g-factor are sensitive to the details of the quantum dot potential. This strong dependence on gate voltages indicates that gate-induced g-tensor modulation[32] contributes to the EDSR mechanism, in addition to direct driving by the SOI.

Measurements of spin blockade in the strong coupling regime highlight a further difference between electron and hole spins. In this coupling regime, hole spin blockade is lifted at finite $B$, leading to a dip around $B = 0$ (Fig. 5 and Supplementary Fig. S6). For electrons, the lifting of spin blockade at finite $B$ is due to mixing of blocked and unblocked spin states by the spin-orbit interaction.[19,26,27] A dramatic signature of this mechanism is strong anisotropy of spin blockade as a function of the angle between $B$ and the nanowire.[19,27] However, in contrast to the case of electrons, here we find that hole spin blockade is almost isotropic at zero detuning and only weakly anisotropic at finite detuning (Fig. 5b and Supplementary Fig. S6). It is important to note that the expected anisotropy is not related to the shape or size of the quantum dot. Rather, it originates from the fixed direction of the Rashba spin-orbit field, as determined by the sample geometry. The absence of spin blockade anisotropy for holes is thus intriguing given the anisotropic p-orbital symmetry of hole Bloch wavefunctions[22], especially since the device geometry is identical to that used for electrons in Ref. [19]. This suggests that either spin-blockade for holes is not described by the same model as for electrons[27] or the dominant spin-orbit interaction for holes is not of the Rashba type.

In summary, our work establishes InSb nanowires as a promising platform for tuneable hole quantum dots and demonstrates the importance of the spin orbit interaction for controlling single hole spins. This provides an avenue for testing the potential for faster and more coherent spin manipulation using holes. In addition, gated InSb nanowires may be a viable platform for Majorana fermion detection using holes[33], with the observed g-factors being sufficiently large to exceed the required g > 2 threshold[34]. Finally, the ability to confine single holes and single electrons in the same nanostructure opens the way for unique applications, such as controlled exchange between hole and electron spins and coupling of distant spins via photons[35], all achievable in III-V semiconductor nanowires using established quantum transport and quantum optics techniques.

**Methods:**

Zincblende InSb nanowires (~100 nm diameter and 2-3 μm long) were grown using metal-organic vapour phase epitaxy (MOVPE) with the [111]-axis oriented along the growth direction.[17] Nanowires were mechanically transferred in air from the growth substrate to a second substrate with pre-patterned gate structures. Contacts were made to selected nanowires. QD measurements were performed in a $He^3$ refrigerator with a base temperature of 300 mK, equipped with a 2-axis vector magnet. Few-hole QDs and spin blockade were observed in two different devices and EDSR was measured on one device in three different cooldowns.

**Acknowledgements:**

We would like to thank L. M. K. Vandersypen and G. Bauer for helpful discussions and comments. This work has been supported by the Netherlands Organization for Scientific Research (NWO), the Dutch Organization for Fundamental Research on Matter (FOM) and the European Research Council (ERC). V. S. P. would like to acknowledge support from NWO through a VENI grant.

**Author contributions:**

V. S. P., S. N., S. M. F., J. W. G. B. and I. W. performed the measurements. V. S. P., S. N., S. M. F. and J. W. G. B. analysed the data. V. S. P., S. N. and J. W. G. B. fabricated the devices. S. R. P. and E. P. A. M. B. provided the nanowires. L. P. K. supervised the project. All authors contributed to writing the manuscript.

**Additional information:**

The authors declare no competing financial interests.

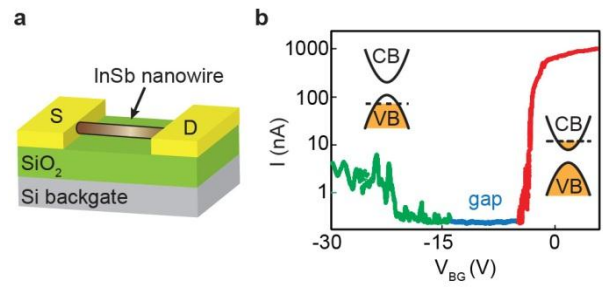

**Figure 1 | Ambipolar transport in an InSb nanowire. a**, Schematic of device used to demonstrate ambipolar transport in InSb nanowires. **b,** Current through an InSb nanowire as a function of the voltage $V_{BG}$ applied to the Si backgate, for a device as shown in **a** (source-drain bias $V_{SD}$ = 10 mV). The nanowire is separated from the backgate by 285 nm of $SiO_x$, and the spacing between the source and drain Ti/Au contacts is ~300 nm. Right (left) insets: when the Fermi level is tuned to the conduction (valence) band, current transport in the nanowire is mediated by electrons (holes). Hole conductance is typically one to two orders of magnitude lower than electron conductance. This could indicate reduced transparency of the nanowire-metal contact interface for holes[16] or lower hole mobility[14]. The measured bandgap is ~0.2 eV, in agreement with the bulk value for InSb[14] (see Supplementary Fig. S1).

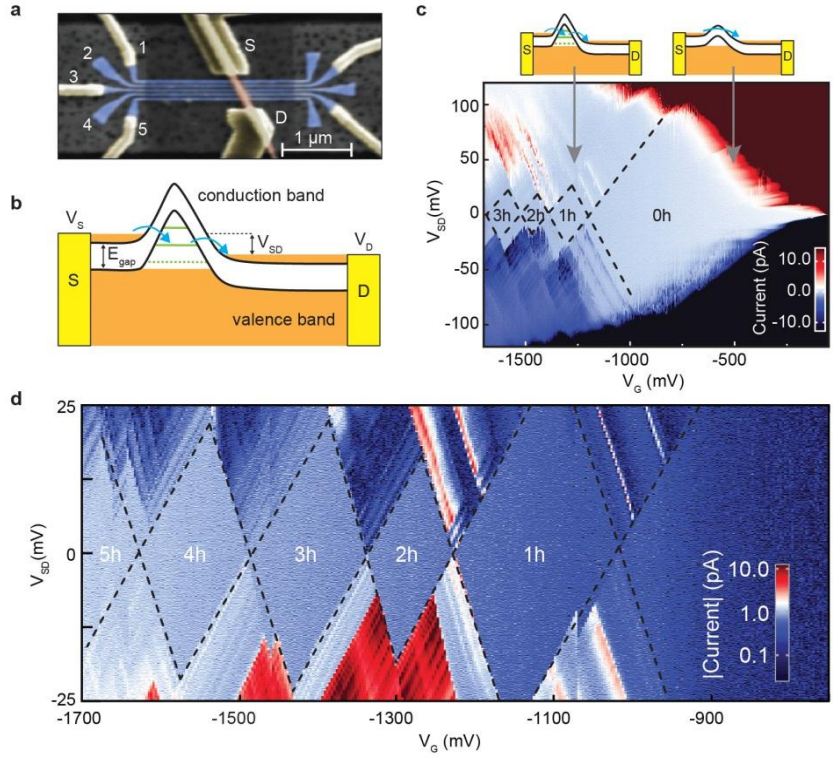

**Figure 2 | Gate tuning between electron transport and hole quantum dot. a,** SEM image of a typical gated InSb nanowire device used for studying QDs. The gate widths and inter-gate spacing are ~30 nm; the source and drain contacts are made of Ti/Al. A wider metallic gate is used to control the charge density in the nanowire segments between the narrow gates and the source and drain electrodes. The wide gate is separated from the narrow gates by a 50 nm layer of $Si_3N_4$; the fine gates are covered by an additional 25 nm layer of $Si_3N_4$. **b,** Schematic of band diagram showing a gate-defined hole quantum dot. Tunneling between the dot and the n-type leads occurs via p-n junctions. **c,** Charge stability diagram (device d1) of a device as in **a**, shown as a function of source-drain bias, $V_{SD}$, and plunger gate voltage, $V_G$. For less negative $V_G$, the nanowire is n-type and current is carried by electrons. For more negative $V_G$, transport is suppressed due to the bandgap of the nanowire. For even more negative $V_G$, a hole QD is formed above the plunger gate; in this regime a finite transport current is observed as a result of tunneling via discrete hole states in the dot. **d,** Hole Coulomb diamonds for device d1 (the dot potential is tuned slightly differently than in **c**). The fluctuating diamond sizes and the absence of any further transitions to the right of the diamond labeled '0h' are consistent with the few-hole regime. However, unambiguous identification of the number of holes would require a charge sensor. From the size of the Coulomb diamonds we estimate that charging energies, $E_c$, of single holes are on the scale of 20 meV and orbital energies, $E_{orb}$, are between 3 and 8 meV.

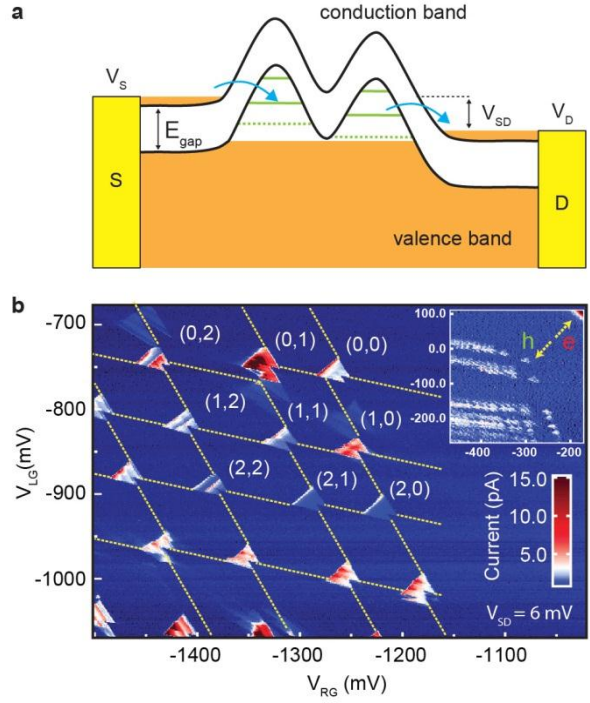

**Figure 3 | Gate-defined few-hole double quantum dot. a,** Schematic band diagram showing a hole double quantum dot and n-type leads. Tunneling onto and off the double-dot occurs via p-n junctions. **b,** Charge stability diagram of a hole double quantum dot as a function of the left and right plunger gate voltages ($V_{LG}$ and $V_{RG}$) (device d2, first cooldown). The larger, dimmer triangles are attributed to an additional QD located in series to the right of the hole double dot and strongly coupled to the drain reservoir (see Supplementary Information). We estimate the size of the dots to be ~15 nm from $l_{dot} \sim \hbar/\sqrt{E_{orb}m_{hole}}$. Inset: stability diagram over a larger area of gate space (device d1, $V_{SD}$ = 12 mV), showing the transition from electron transport (upper-right corner) to the hole double dot regime (lower-left region).

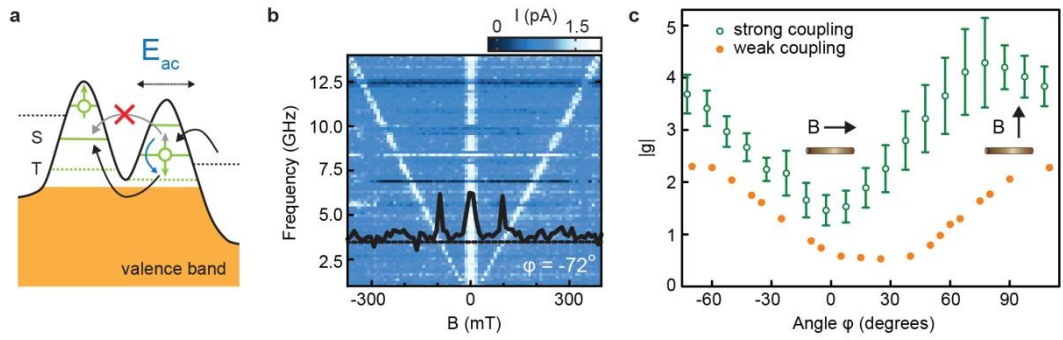

**Figure 4 | Electric-dipole spin resonance and hole g-factor anisotropy. a,** Interdot tunnelling is suppressed by spin blockade whenever unpaired holes in each dot form a triplet state (T). A microwave-frequency electric field of amplitude $E_{ac}$ applied to the right plunger gate induces spin rotation to a singlet (S) by means of EDSR, lifting the spin blockade. **b,** EDSR for weak interdot coupling (device d2, second cooldown). The line cut is at $f = 3.4$ GHz. In addition to the EDSR resonances, we observe a lifting of the spin-blockade near $B = 0$, attributed to the hyperfine interaction. **c,** Anisotropy of the hole g-factor extracted from EDSR measurements for different angles $\varphi$ between the applied magnetic field $B$ and the nanowire axis in the plane of the sample surface. Weak coupling g-factor data is from the second cooldown of device d2 and strong coupling data from the third cooldown of the same device.

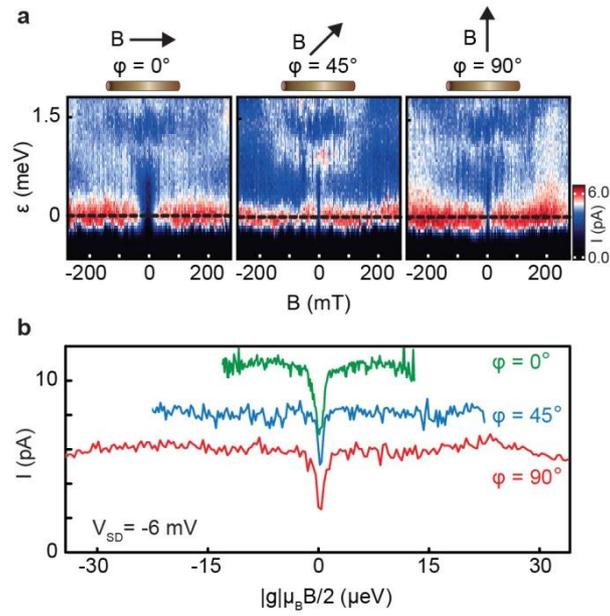

**Figure 5 | Hole spin blockade for strong interdot coupling. a,** Double-dot current vs. detuning ε and magnetic field $B$ in the strong-coupling regime for three different angles (device d2, third cooldown, $V_{SD}$ = -6mV). Spin blockade is observed as a dip near $B$ = 0. **b,** Cuts along the dotted lines at ε = 0 in **a**. The applied magnetic field is scaled by the effective g-factors at each angle (see Fig. 4). For clarity, the data for $\varphi$ = 45° and $\varphi$ = 0° are offset vertically by 2.5 pA and 5.0 pA, respectively. Offsets of several mT due to the magnet were subtracted from the data in **a** and **b**.

# Supplementary Information

# Electrical control over single hole spins in nanowire quantum dots

*V. S. Pribiag, S. Nadj-Perge, S. M. Frolov, J. W. G. van den Berg, I. van Weperen., S. R. Plissard,*

*E. P. A. M. Bakkers and L. P. Kouwenhoven*

**Contents:**

1. **Bandgap of InSb nanowires**

We extract the bandgap of our InSb nanowires using the basic device described in Fig. 1 of the main text. On the left side of Fig. S1, the Fermi level is in the valence band, resulting in a finite current away from zero bias. At less negative $V_{BG}$, the Fermi level is inside the bandgap and the current is suppressed. On the right side of the figure, at even less negative $V_{BG}$, current is restored as the Fermi level moves into the conduction band. We extract the bandgap, ~0.2 eV, from the extent of the non-conducting region, as shown by the arrow in Fig. S1. This value is in agreement with the gap of bulk InSb (~0.17 eV at room temperature and ~0.23 eV at low temperatures [1]), and is confirmed by similar measurements in one other InSb nanowire device.

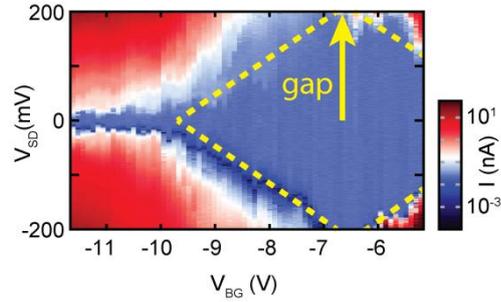

*Fig. S1: Current through an InSb nanowire as a function of $V_{SD}$ and $V_{BG}$.*

## 2. Origin of the background triangles in Fig. 3b:

The additional triangles in the background are consistent with a single quantum dot located in series to the right of the main hole double dot (see figure below). In order not to obscure the main features, this single dot must be strongly coupled to the drain reservoir. Tunnelling between the right hole dot (RD) and this extra dot (ED) leads to the double dot features in the background. This specific spatial configuration is supported by the fact that the double dot associated with the background triangles couples as strongly to $V_{RG}$ and less strongly to $V_{LG}$ than the hole double dot associated with the main triangles. It also explains why the extra triangles are so dim: they are located in gate space regions where the left hole dot (LD) is in Coulomb blockade, meaning that direct tunnelling via LD is suppressed and charge transport is only possible via co-tunnelling [2]. The strong coupling to the drain broadens the levels of ED, as evidenced by the smooth baselines of the background triangles and the uniform current within [3]. The somewhat sharper resonances on the lower sides of two of the background triangles are consistent with resonant tunnelling via the excited states of RD, which are not lifetime broadened. We emphasize that our results are not affected by this dot, since no background triangles were observed near the charge transitions where the EDSR spectra and spin blockade were measured (see Fig. S4b and Fig. S5a).

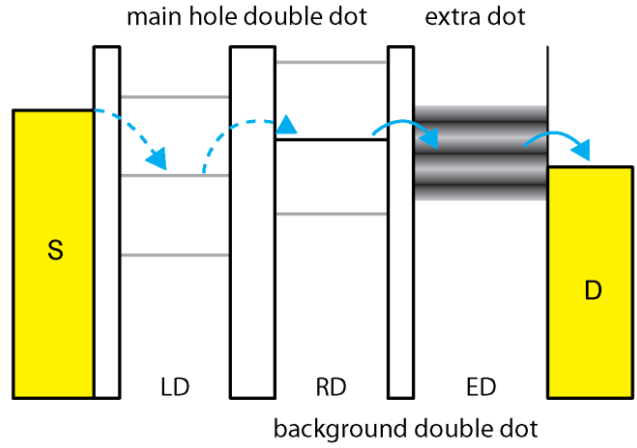

*Fig. S2: Schematic of transport through the background triangles in Fig. 3b. The main triangles in Fig. 3b are associated with the main hole double dot. The fainter background triangles are consistent with tunnelling between RD and ED. ED is coupled to the drain through a ~100 kΩ Schottky tunnel barrier formed at the interface between the nanowire and the metallic contacts.*

## 3. Zeeman splitting of QD lead resonances

Transitions to excited states of the dot lead to resonances in the charge stability diagram of a QD.[4] Since an empty dot has no excited states, no resonances due to excited states can terminate at the N=0 diamond. To investigate the origin of these resonances (Fig. S2a) we rely on the very different magnetic field dependence of holes and electrons in our InSb nanowire devices. Fig. S2b shows magnetic field splitting of two resonances ending at the N=0 diamond. The splitting corresponds to a g-factor of ~35, in agreement with values for the electron spin, and much larger than the g-factor of holes extracted from EDSR measurements (Fig. 4 in main text). This suggests that these resonances are likely due to peaks in the electron density of states in the n-type leads [5] and not to hole excited states in the QD.

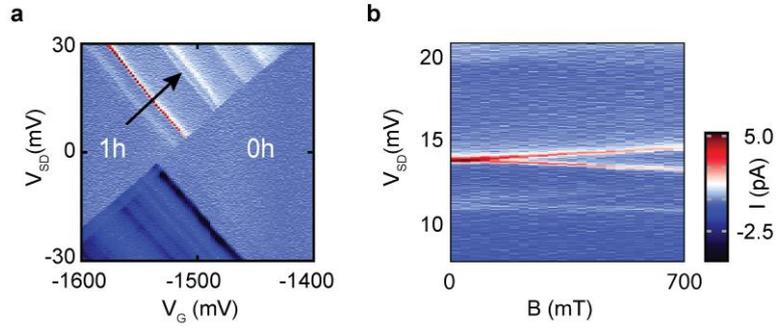

*Fig. S3: Zeeman splitting of QD lead resonances.* **a,** *Stability diagram for device d1, showing the transition between the 1-hole and the 0-hole configurations.* **b,** *Magnetic field dependence of the resonances along the cut indicated by the arrow in* **a**.

## 4. Spin blockade data in the weak coupling regime

Fig. S4a shows a hole double dot under positive source-drain bias. A hole tunnelling from the right lead onto the right dot can form a (1,1) triplet with the hole in the left dot. Whenever this occurs tunnelling to the (2,0) singlet is forbidden by spin conservation, resulting in a suppressed current. Fig. S4b shows an example of an interdot transition (for weak interdot coupling) that exhibits spin-blockade. Here, the number of holes in the double dot is ~10. We identify spin blockade by magnetic field dependence and bias asymmetry. For positive bias spin blockade suppresses the tunnelling current for detuning values between 0 and ~0.9 meV (Fig. S4c). The spin blockade is lifted near $B = 0$ as a result of spin mixing by the hyperfine interaction.[6-11] For negative bias no current suppression is observed, as expected for spin blockade. This occurs since spin exchange with the left lead ensures that the dot can always be initialized in the (2,0) singlet for negative bias and from this state one of the holes can always tunnel to the right lead via the (1,1) singlet, leading to a finite transport current.

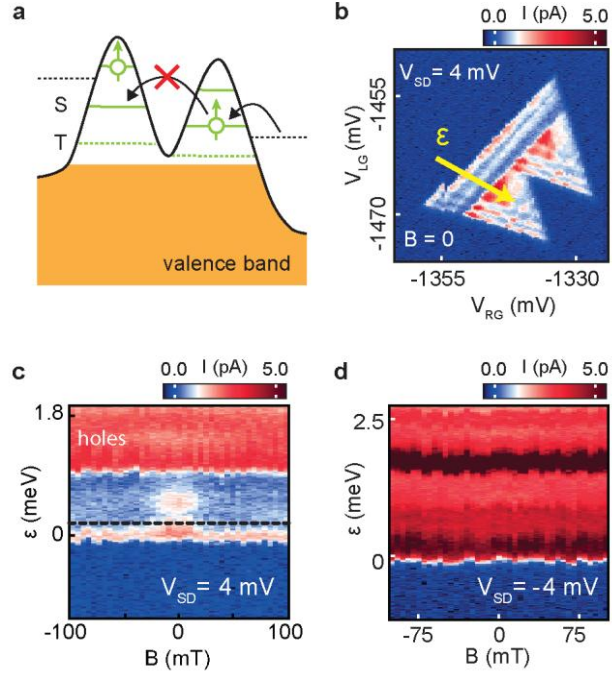

*Fig. S4: Spin blockade data for weak coupling regime of a hole double dot.* **a,** *Schematic of spin blockade for hole double dot.* **b,** *Charge stability diagram near a spin-blocked transition, in the weak-coupling regime (device d2, second cooldown). The arrow indicates the detuning axis used in* **c** *and* **d**. **c,** *Double dot current vs. detuning ε and applied magnetic field B for positive $V_{SD}$.* **d,** *Double-dot current vs. ε and B for negative $V_{SD}$.*

## 5. Comparison of hyperfine coupling strength between holes and electrons

To compare the hyperfine interaction strengths of holes and electrons, we rely on the standard method from Refs. [6] and [7], which relates the width of the hyperfine-induced $B=0$ peak in the spin blockade regime to the RMS fluctuations of the hyperfine field, $B_N$. By applying this method to the data in Fig. S5c, we obtain $B_{N,e}$ ~1 mT for electrons. This corresponds to an energy scale $E_{N,e} = g\mu_B B_N$ ~2 µeV. The same method, applied to the hole double-dot regime of the same device (see Fig. S5d), yields $B_{N,h}$ ~6 mT and $E_{N,h}$ ~0.8 µeV. The ratio of the magnitudes of the effective hyperfine interaction strengths of electrons ($A_e$) and holes ($A_h$) is given by $A_e/A_h = (E_{N,e} * \sqrt{N_e})/(E_{N,h} * \sqrt{N_h})$, where $N_e$ and $N_h$ are the number of spinful nuclei in the electron and hole dots, respectively. [4] The hole dots are about half the size of the electron dots (~1/8 times the volume, see main text), thus yielding $A_e/A_h$ ~7. This is, to our knowledge, the first estimate of the strength of hyperfine interaction of holes in gate-defined quantum dots. This sevenfold reduction in the hyperfine interaction strength for holes is close to the values obtained by optical measurements on self-assembled dots. [12]

We note that this standard method of extracting the hyperfine coupling in transport quantum dots, may sometimes overestimate $B_N$.[9, 13] However, we expect that this has no effect on the ratio $A_e/A_h$, since both $B_{N,e}$ and $B_{N,h}$ were obtained from the same device using the same setup.

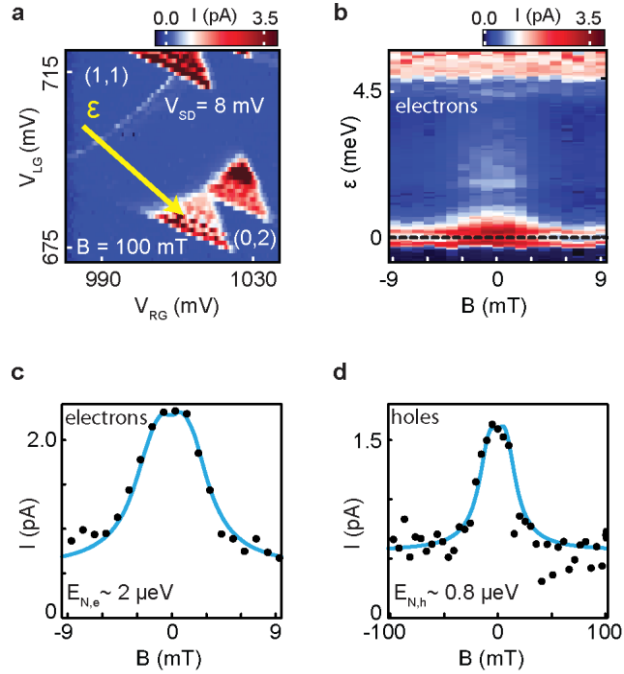

*Fig. S5: Spin blockade data for weak coupling regime of an electron double dot and comparison of hyperfine peak widths.* **a,** *Charge stability diagram near a spin-blocked transition of an electron double dot, in the weak-coupling regime (device d2, first cooldown). The arrow indicates the detuning axis used in* **b**. **b,** *Double-dot current vs. detuning ε and applied magnetic field B for positive $V_{SD}$. An offset of several mT due to the magnet was subtracted from the data.* **c,** *Cut at fixed ε for along the dashed line in* **b**. *The line is a fit to the data using the model of Ref. [7], which yields $E_{N,e}$ ~2 µeV.* **d,** *Cut at fixed ε along the dashed line in Fig. S4c. The line is a fit to the data using the model of Ref. [7], which yields $E_{N,h}$ ~0.8 µeV.*

## 6. Additional measurements of the angular dependence of spin blockade in the strong coupling regime

Fig. S6 shows detuning vs. magnetic field data for the spin-blocked transition (1,1)-(0,2) in the strong interdot coupling regime (device 2, first cooldown). As the magnetic field is rotated with respect to the nanowire axis, no significant anisotropy of the spin blockade leakage current is observed (the leakage current varies by a factor of only about two as function of angle). This is consistent with the data from Fig. 5, which was measured on a different transition and in a different cooldown. We conclude that the absence of significant anisotropy of the leakage current is a robust feature of the hole spin blockade. In contrast, clear anisotropy of the leakage current (a change by an order of magnitude) has been observed for electron spin blockade in the same device (cf. Figs. 4(e) and 4(i) in Ref. [13]). This anisotropy was attributed to the fixed direction of the Rashba spin-orbit field vector for the double dot geometry. The absence of strong anisotropy for hole spin blockade may point to different angular dependence of the SOI or a different spin-blockade lifting mechanism than for electrons.

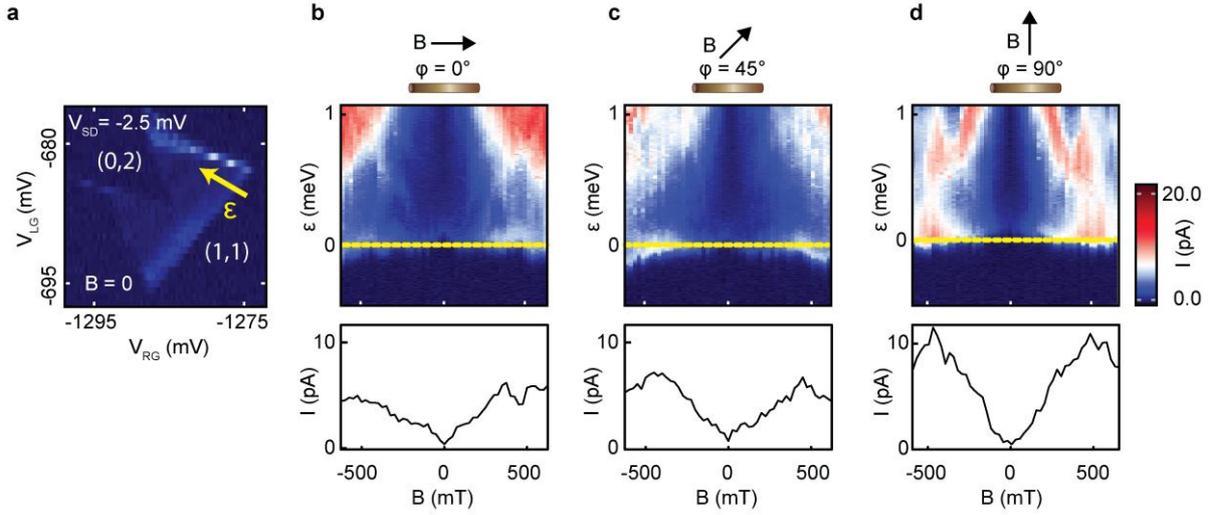

*Fig. S6: Hole spin blockade for strong interdot coupling.* **a,** *Charge stability diagram near the (1,1)-(0,2) transition, in the strong-coupling regime (device d2, first cooldown, $V_{SD}$ = -2.5mV). The arrow indicates the detuning axis used in **b-d**. **b-d,** Double-dot current vs. detuning ε and magnetic field B in the strong-coupling regime for three different angles. Spin blockade is observed as a current dip near B = 0. An offset of several mT due to the magnet was subtracted from the data in **d**.*

## 7. Method for extracting the g-factors

The g-factors were extracted from EDSR measurements by monitoring the spin blockade leakage current as a function of applied microwave frequency, $f_0$, and magnetic field value, $B$. Whenever the spin resonance condition is fulfilled, the spin blockade is lifted, resulting in increased current through the dot, as illustrated in Figure 4a. For weak interdot coupling, we swept $f_0$ for a fixed value of $B = 270$ mT, repeating the measurement for different angles between the field and the nanowire axis. The g-factors were then calculated from these data using $f_0 = g\mu_B B/h$. For strong interdot coupling we swept $f_0$ vs. B, obtaining data of the type shown in Figure 4b for each angle. The g-factors were calculated from the slope of the resonance lines using the same equation. The larger error bars for strong coupling g-factors are due to the lower contrast of the EDSR resonance in this regime. This occurs as a result of partial lifting of spin blockade at finite $B$ [13] (see e.g. Fig. 5 and Fig. S6).